# SAFETY ANALYSIS METHODS FOR COMPLEX SYSTEMS IN AVIATION

**Ítalo Romani de Oliveira, José Alexandre T. Guerreiro Fregnani, Gláucia Costa Balvedi, Michael L. Ulrey, Jeffery D. Musiak**
*Boeing Research & Technology*
italo.romanideoliveira@boeing.com, jose.a.fregnani@boeing.com, glaucia.c.balvedi@boeing.com, michael.l.ulrey@boeing.com, jeffery.d.musiak@boeing.com

**Ricardo Alexandre Veiga Gimenes, João Batista Camargo Jr., Jorge Rady de Almeida Junior**
*Universidade de São Paulo – Escola Politécnica da USP*
*Grupo de Análise de Segurança – GAS*
rveiga@usp.br, joaocamargo@usp.br, jorgerady@usp.br

## ABSTRACT

Each new concept of operation and equipment generation in aviation becomes more automated, integrated and interconnected. In the case of Unmanned Aircraft Systems (UAS), this evolution allows drastically decreasing aircraft weight and operational cost, but these benefits are also realized in highly automated manned aircraft and ground Air Traffic Control (ATC) systems. The downside of these advances is overwhelmingly more complex software and hardware, making it harder to identify potential failure paths. Although there are mandatory certification processes based on broadly accepted standards, such as ARP4754 and its family, ESARR 4 and others, these standards do not allow proof or disproof of safety of disruptive technology changes, such as GBAS Precision Approaches, Autonomous UAS, aircraft self-separation and others. In order to leverage the introduction of such concepts, it is necessary to develop solid knowledge on the foundations of safety in complex systems and use this knowledge to elaborate sound demonstrations of either safety or unsafety of new system designs. These demonstrations at early design stages will help reducing costs both on development of new technology as well as reducing the risk of such technology causing accidents when in use.

This paper presents some safety analysis methods which are not in the industry standards but which we identify as having benefits for analyzing safety of advanced technological concepts in aviation.

# 1. INTRODUCTION

The fear of the obscure risks hidden in complex technology are not a novelty. Several world famous films express this idea by showing human-made robots or personified systems rebelling against their creators. Despite the fantasy, it could be argued that the real versions of the same stories are the accidents involving automated functions. When it is possible to distinguish functions necessary to a system to operate, fact which is not universally valid, the usual development path is to have some of these functions performed by humans and some functions performed by what we may generically call "automation". In highly complex systems, it may well happen that functions overlap each other or that, due to the needs and limitations of the development process, functions are not or even cannot be well-defined. Among these cases is the Air Transportation System, which in the primordia was a set of functionally independent elements, where the main integrator agent was the human and each flight was unique due to the intense labor necessary for its safe execution. In contrast, the present reality of Air Transportation is a System of Systems, where heavy lift tasks, including communication and integration, are largely automated, and the human can concentrate his/her efforts in supervisory tasks. But there is still much more evolution to come and this paper is an attempt to present methods and research work which can help clarify whether or not a technology or a Concept of Operations can achieve the expected safety goals.

# 2. CURRENT STATUS OF SAFETY REGULATIONS IN AVIATION

For the current status quo, there is a strong body of mandatory standards for development of aviation systems, either airborne (avionics), in the space (Communication, Navigation and, recently, Surveillance, thus CNS) or on the ground (Air Traffic Management – ATM). For avionics systems, the chief standards are ARP4754 / ED-79A [1], for CNS/ATM systems there are DO-278 / ED-109 [2], the latter ones being subordinated to the former ones, in the following sense: ARP4754 / ED-79A lay down the complete set of principles on which a certification process must be based, including the need of a safety analysis process on the system design, and the principles on which a system development process must be based in order to ensure the safety goals (and, indirectly, quality levels). These standards originated in the aviation industry in a bottom-up fashion and later on were adopted by the regulatory authorities. They do not describe how to establish quantitative Target Levels of Safety (TLS) on the application operational level, since this is a societal decision based on risk acceptance and, generally, is defined by governmental bodies, such as ICAO, FAA and Eurocontrol.

Actually, regulatory authorities use a mix of qualitative and quantitative requirements. ICAO Doc 9859 [3] presents requirements for *Safety Management* but does not define quantitative TLS. It states "*whenever quantitative safety performance targets are set, it must be possible to measure, or estimate, the achieved level of safety in quantitative terms. Use of quantitative data helps clarify most decisions and should be used where available*". For the specific aspect of collision risk in air traffic, ICAO's Annex 11 [4] establishes a quantitative TLS for the collision probability. It is known that this TLS is partly based on historical statistics, partly on a "desirable" target. Besides, it equitably splits the collision risk per each of the three spatial dimensions (lateral, longitudinal and vertical), resulting in the number $5 \times 10^{-9}$ per dimension. This equitable and independent splitting is assumed but not rationally justified. For illustration, one can consider the case where one aircraft is at the wrong altitude and the other is at the wrong track. If a collision happens in this case, to which dimension would the event be computed in the statistic? This and other questions motivated a critical analysis of the dominant methods of safety assessment, which we present below.

## 2.1. Dominant approaches of Safety Assessment and their drawbacks

Whenever there is a new Concept of Operations (ConOps) or a new design in aviation, or an incremental change to an existing ConOps or design, the competent aviation authority requires a demonstration of safety for it, what we may generically call "Safety Assessment". As mentioned above, the Safety Assessment has logical and quantitative arguments. The use of the term *qualitative* is avoided here because we think is poor of sense. The elaboration of arguments of each of these types is achieved by a dominant methodology: the logical arguments are stereotyped by the "Safety Case" and the quantitative arguments are stereotyped by the "Bow-Tie Model".

### 2.1.1. The Safety Case approach

"A Safety Case is a structured argument, supported by evidence, intended to justify that a system is acceptably safe for a specific application in a specific operating environment" [5]. Basically, a safety case builds a rational thesis demonstrating whether or not the system conceptual design is safe, supported by a set of evidences about the system and the system environment. In order to standardize and facilitate this argument building, Goal Structuring Notation (GSN) is being increasingly more used [6, 7], but there is not a uniformly expected format for a safety case. Such one may contain or refer to quantitative analyses, but not as a pre-requisite. Safety cases are used as acceptable means of compliance when a regulatory authority does not have a repeatable and prescribed process for compliance assurance or certificate issuance, or when the system under scrutiny is a kind of its own and no existing standard would be enough.

### 2.1.2. Prescriptive approaches

On the other hand, when it is assumed that the system being designed is similar in nature to a series of other existing systems (e.g. a new aircraft design is similar to existing designs in far too many aspects), then it is possible to standardize the means of compliance based on certain assumption about the system's features. For example, the avionics software development process has to conform to DO-178C [8] and this makes its verification, validation and documentation highly standardized; since the invention of fly-by-wire, the principal avionics functions haven't significantly changed, such that there is a common classification for their criticality levels and, regarding the hardware, compliance with the hardware development standard DO-254 [9] is the required means of compliance. These standards are complementary to the FAR airworthiness regulations, which require compliance by performance parameterization, test criteria, etc. This approach also relies on official examiners which provide appraisals of compliance which can be verified for coherence against the regulations. This approach, which is named here as *prescriptive*, exempts the pleading party of elaborating a formal safety case because the proofs required by the applicable standards are sufficient means of compliance.

The prescriptive safety assessment approaches in the industry, as we observe, almost invariably requires the defendants to present a quantitative safety verification. It can be said that, because the prescriptive approaches promote a divide-and-conquer strategy, it is easier to break down the system in functions and components and, therefore, it is possible to identify basic failure events, know how these events trigger hazards, and which consequences a hazard may have. Most importantly, the possibility of quantifying the probability of occurrence of basic failure events and understanding how these probabilities influence the probabilities of intermediate events and final consequences leaves an opportunity for compliance verification which regulatory authorities should not dismiss. And here is when the bow-tie model enters into action.

### 2.1.3. The Bow-tie model

An example of the Bow-tie model is illustrated in Figure 1. Such a model links the causes of a hazard (modelled using Fault Tree Analysis (FTA)) and the consequences of a hazard (modelled using Event Tree Analysis (ETA)). The point in the Fault Tree (FT) hierarchy at which the link to an Event Tree (ET) is established is known as a pivotal event, which is commonly found to correspond exactly to what is called a *hazard*. The pivotal events typically correspond with the main system and or subsystem hazards. One FT/ET pair is constructed for each hazard and values are ascribed both to the probability of occurrence of each casual factor in the FTs and to the probability of success or failure of the outcome mitigations represented by the branches of the ETs. Using the facilities of a mature FTA / ETA tool, the overall probability of an accident from all causes can be determined and compared to the safety target(s).

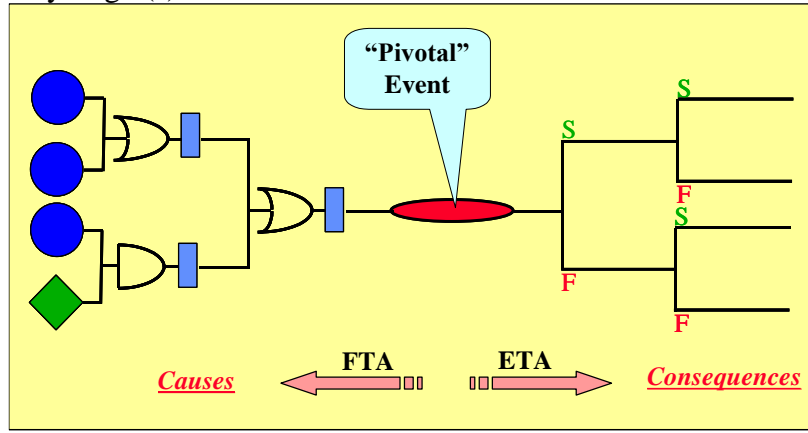

**Figure 1: a generic example of Bow-tie model (Source: [10]).**

It is interesting to observe that all major regulatory standards in safety critical industries recur to the bow-tie model either explicitly or implicitly and, besides, they are similar in their schemes to define safety targets [11]. These are based on a risk classification table where, depending on the severity of the consequence of a hazard or failure event, a maximum probability of occurrence is imposed to its occurrence. This is true for a widely embracing list of standards:

- ARP4761 [12], which is the foundation for safety assessment supporting ARP4754 / ED-79A and their subordinated standards, applicable for avionics systems and ground Communication, Navigation and Surveillance (CNS) systems; here the safety target is referred to as "Development Assurance Level" (DAL);
- IEC 61508 [13], used in a variety of safety critical applications, either directly or indirectly, by means of derived industry-specific standards; here the safety target is expressed as Safety Integrity Level (SIL);
- Eurocontrol's Safety Assessment Methodology (SAM) [14], which supports ESARR 4 [15], a regulation for Risk Assessment and Mitigation in ATM.
- DoD's Mil-Std 882C [16] and MoD's Def Stan 00–56 [5], aimed at military systems.

Despite this list being neither large nor exhaustive, it embraces civil aviation, military systems, automotive, rail, nuclear, process industries and more.

### 2.2. Drawbacks of the dominant approaches

The reliance of safety cases on GSN for argument building helps achieving coherence and clearness in the argumentation logic, but does not guarantee full consistency among the several safety case sections and individual elements. Even for small argument trees, it is unfeasible to check every node against every node, not to mention transitive implications among nodes. From a philosophical point of view, it can be said that safety are susceptible *narrative fallacies*. Such term was coined by Nassim Taleb as a concept that "*addresses our limited ability to look at sequences of facts without weaving an explanation into them, or, equivalently, forcing a logical link, an arrow of relationship upon them. Explanations bind facts together. They make them all the more easily remembered; they help them make more sense. Where this propensity can go wrong is when it increases our impression of understanding*" [17].

If narrative fallacies sound too philosophical in this engineering context, other very compelling arguments are provided by the Natural Accident Theory (NAT) of Charles Perrow. In his work [18], he presents numerous and detailed examples, such as: Three Mile Island (TMI), aircraft crashes, marine accidents, dams, nuclear weapons, and DNA research. Through detailed reconstruction of events, he shows that the accidents are not only "unexpected, but are *incomprehensible* ..." to those responsible for the safe operation of those systems. "In part, this is because in these human-machine systems the interactions literally cannot be seen." For these reasons, Perrow considers "normal accidents" and "system accidents" to be synonymous.

Both the narrative fallacy and NAT arise and grow with system complexity, as well as their effects are proportional to this complexity. The problem is that the industries cannot revert the technological advances which happened for good reasons. Despite the complexity increase, accident rate in aviation has consistently dropped in the last decades. It is inconceivable a new commercial airplane design without fly-by-wire, for example. These advances allowed for efficiency gains, by decreasing aircraft weight and maintenance costs, among other benefits. Thus, complexity is unavoidable, and a corresponding evolution in safety analysis methods, either by enhancing safety case methods or complementing them with diverse methods, is needed.

The multi-aspect prescriptive methods are more robust to the complexity-borne problems just pointed out, mostly because they use practice-oriented criteria and checkpoints, however they are not infallible. Firstly, because testing complex systems can never be exhaustive and, secondly, because the bow-tie model is also susceptible to narrative fallacies, NAT. Thirdly, one of its central elements, the bow-tie model, has a number of drawbacks.

Despite the bow-tie model being conceptually simple and easy to build, it has several hidden assumptions and drawbacks. The most fundamental hidden assumption is that the list of hazards is complete or, at least, that no significant hazard is omitted. This cannot be guaranteed and often not taken into account and, despite there is a large variety of structured methods for eliciting hazards [19, 20], it is very hard to guarantee their completeness and consistency. Often, hazards are dependent on each other, and their dependency relation is hard to quantify.

A bow-tie model also becomes problematic when multiple events which cause hazards have themselves common causes and thus are dependent in the probabilistic sense. In these cases, which are commonplace, special software tools are needed to adequately build the fault tree side and perform the probability calculations [21, 22], thus there must be some way to check that these tools can be relied upon. Furthermore, on the consequence side of the bow-tie, the complication happens when multiple consequences of an event are neither mutually exclusive nor concomitant. For example, an aircraft in the middle of a RNP-X approach may lose this capability. This event can be considered a hazard and might have as consequence some subsequent flight instability, a TCAS advisory or both, and it is hard to access how much one influences the other.

**2.3. Auxiliary methods**

Of course, GSN and the Bow-tie Model are not the only tools for safety analysis and some of their drawbacks can be mitigated by complementation or substitution by other methods. Below are some methods with more specific use, which can be found as reference in the industry standards cited above.

**2.3.1. Markov Analysis**

Markov analysis consists in modelling a system or subsystem as a Markov chain composed by states and state transitions, from which it is possible to calculate the probability of the system being at some state when observed at a random time. It is very useful for modelling fault tolerant systems, which may recover from faults either by maintenance or system reconfiguration, however it becomes difficult to use for systems with large number of states. Thus, in general, it is used for generating failure probabilities at subsystems which feed the bow-tie model.

### 2.3.2. Failure Mode and Effects Analysis (FMEA)

According with ARP4761, FMEA is "*systematic, bottom-up method of identifying the failure modes of a system, item, or function and determining the effects on the next higher level. It may be performed at any level within the system (e.g., piece-part, function, blackbox, etc.). Software can also be analyzed qualitatively using a functional FMEA approach. Typically, an FMEA is used to address failure effects resulting from single failures*". So, the latter sentence itself points to the main limitation of this method: it has a tunnel vision approach and, as so, it should not be used directly to add up probability in a verification against a safety target if the unsafe event defining that target is influenced by events which are not properly considered in the FMEA. Saying it in a different manner, an effect item in an FMEA item could be reached by different paths and the FMEA technique itself does not offer elements to handle this combinatorics. Therefore, FMEA should be used in conjunction with higher level methods such as the bow-tie model or some of the other ones described in this paper.

### 2.3.3. Formal Methods

Formal methods use principles of mathematical reasoning for allowing a "correct-by-construction" system design. According with IEC 61508, Part 7, the aim of formal methods is to "*transfer the principles of mathematical reasoning to the specification and implementation of technical systems, therefore increasing the completeness, consistency or correctness of a specification or implementation*".

The process of using formal methods for safety assurance is roughly composed by the following steps:
1) Elaborate the system model using a formal language;
2) Elaborate the desired safety properties using a formal language compatible with the language of the formal model; and
3) Use a formal reasoning method to check whether or not the system model upholds the safety properties.

Thus, formal methods are used to answer yes or no questions about the system safety, or to restrict the specification in order to maintain the safety properties. They are not used for quantitative analyses. The advantage of using formal methods is to decrease testing effort and, more than that, achieve coverage of system states that would not be feasible to achieve with testing. Its relevance for the safety critical systems community has been steadily growing as it is confirmed by their inclusion as an alternative means for system development in the avionics software standard DO-178C [8], in conjunction with the issuance of its supplement DO-333 [23]. There are several types and flavors of formal methods, and a thorough review of this topic can be found in [24].

Despite formal methods can be powerful if properly used, they also have drawbacks. The first problem regards the modeling power of the formal languages used. It often happens that these languages are not capable of representing the desired engineering system in a model, more evidently when if the system has non-linear dynamics. A similar limitation happens with the specification of safety properties. Sometimes, the system representation is possible, but the verification algorithm does not terminate or terminates in impracticable time. Finally, there is the fact that using these techniques require mathematical and computational skills which are difficult to master.

### 2.4. Wrapping up the critics to the status quo of safety analysis methods

The critics in this section are not meant to advocate the abandonment of the current approaches, which have the merits of helping the industry to provide us with systems on which we trust for transportation, energy, chemicals, etc. However, the fact that they become more cumbersome and prone to errors as the system complexity grows has the effect of slowing down technological innovation due to the difficulty of elaborating sound safety analyses without reference to known designs and risks. This is a motivation for seeking complementary approaches that can provide reliable safety verifications at early design stages of innovative complex systems.

## 3. NON-STANDARD SAFETY ANALYSIS METHODS FOR COMPLEX SYSTEMS

This section presents some safety analysis methods which are not part of the industry standards, but which have been demonstrated as successful in many practical safety studies of complex systems.

### 3.1. Multi-Agent Dynamic Risk Models (MA-DRM)

Heterogeneous engineering systems having both technical infrastructure such as hardware and social infrastructure such as agents and institutions are generally referred to as socio-technical systems. By way of example, air traffic management and power network management systems include a close interplay of diverse technical artifacts and social organizations. A human individual operating such systems is driven both by her or his internal states and processes and by external inputs, inputs which may exist previously to the operation itself. Such an individual interacts with the system and with other individuals related to the system, hence the higher level system composed by the human agents and the technological system is named *socio-technical system*. A good reference on modelling socio-technical systems can be found in [25].

Multi-agent dynamic risk models (MA-DRM) have been proven successful for analyzing complex socio-technical systems concerning safety properties [26, 27]. This methodology combines concepts of distributed artificial intelligence with stochastic estimation methods. Within the greater area of distributed artificial intelligence, it uses the so called Multi-Agent systems [28] and, from the stochastic estimation methods, it needs a stochastic representation of the system which can be used by a Monte Carlo-based algorithm for evaluation of event probabilities. The stochastic estimation method which has demonstrated great adherence for this type of application is TOPAZ [29, 30]. Indeed, in [27] it is demonstrated that, in a safety assessment of a runway crossing operation, MA-DRM helped to identify hazardous sequences of events which were not identified when using bow-tie models in a standard analysis, and the MA-DRM stochastic simulation-based approach resulted in higher probabilities of incident and accident events.

### 3.2. STAMP (Systems-Theoretic Accident Model and Processes)

Even though engineering processes vary among organizations, every sensible engineering effort shares some similar phases. System safety engineering begins with finding hazards in the system and emerging constraints on the system design aiming to mitigate these hazards. System safety continues by ensuring that modules are designed in such a way as to enforce the safety constraints. Accidents occur when a safety constraint is not adequately enforced by the system's components [31].

The basics of System-Theoretic Accident Model and Processes (STAMP) is to identify leading indicators for risk management based on the assumptions underlying our safety engineering practices and on the vulnerability of those assumptions rather than on likelihood of loss events [32]. STAMP is a qualitative and comprehensive accident model created by Dr. Nancy Leveson to analyze accidents in systems [33]. The main concepts in STAMP are constraints, control loops and process models, and levels of control. The most important concept in the new model is not an event, but a constraint. STAMP considers the interaction among human, hardware, and software, and the hypothesis underlying STAMP model is that the system theory is a useful way to analyze system accidents. This notion of safety implies that accidents occur when outer disturbances, component failures, or no desirable interactions among system components are not correctly handled by the control system. Therefore, they result from inadequate control of safety-related constraints on the development and operation of the system.

The cause of an accident may be viewed as the result of a lack of constraints forced on the system design and on operations, that is, inadequate constraints enforcement on behaviors at each level of a system. Safety can be understood as a property that arises (emergent property) when the system components interact with an environment. Emergent properties are measured by a set of constraints related to the behavior of the system components. Accidents result from a lack of constraints set on the interactions [33]. As an overview, STAMP assists in recognizing scenarios, non-functional interactions and the incorrect models and processes, which will be used in development for a safer system.

### 3.3. Other methods

Both methods above, MA-DRM and STAMP, provide means to end-to-end analysis of models of socio-technical systems, however they use very different approaches, that is one of the reasons why they were dedicated to exclusive sections in this paper. The first one requires rigorous mathematical tools and analysis, while the second is more purely argumentative. Each one has advantages and disadvantages, and we believe that they offer complementary views of safety. Anyway, there are many other methods which deserve consideration from safety analysts, but we could not be exhaustive in this paper. For a more complete research on safety assessment methods applicable to aviation, we recommend [34].

## 4. APPLICATIONS DEMANDING NON-STANDARD SAFETY ANALYSES

Several new technologies in aviation are expected to become standard sometime in the future, with some of them being established earlier and some later but, in our view, they require the use of the non-standard methods of Section 3 in order to clarify to the safety authorities which safety targets they can meet.

### 4.1. GBAS GNSS precision landings

The use of differential GPS was first developed to mitigate the effects of the selective availability (SA) in the signal that could refrain operations from precision approaches, with meters of accuracy. The Large Area Augmentation Systems (LAAS) was therefore developed by FAA in early 90's as a suitable solution using the DGPS concept which provides higher lateral accuracy via position corrections and integrity alerts transmitted via VHF data link. Successful tests influenced ICAO to nominate the augmented GPS as the future air navigation precision approach landing system in 1995, when requirements for Civil Operations of DGPS systems, later called GBAS (Ground Based Augmentation System), were then included in the Annex 10 [35]. This precludes accuracy, integrity, integrity risk, continuity, continuity risk and availability recommended limits to be used by the States on GBAS certification regulations. These parameters are embedded in the on-board navigation systems (implemented as monitoring and protection algorithms) in such way that it is possible for pilots to determine wherever or not is safe to conduct the approach. Worth to mention that for GBAS certification, the integrity dominates the accuracy requirements (therefore the verification of integrity also verifies the accuracy). Also, continuity is verified on probabilities linked to fixed levels of safety [36].

The developments of this technology brought questions on how to protect users against internal system failures and external agents. The primary system failure mode is obviously related to ground receivers out (in total of four). However the main challenges of failure assessments are related to the impact on the integrity of the position solution and availability of services as follows: loss of synchronization of ranging measurements (such as clock-runoff), space weather effects (such as ionospheric scintillations, very predominant on South Atlantic and over Brazilian territory), constellation weakness (related to geometric dilution of precision and number of satellites available), and radio-frequency interference.

Some of the complexities arising from these scenarios are the switching to contingent procedures in case the integrity alert kicks in, which has cognitive challenges for the operators, which also may lose confidence in the navigation system as a whole, or progressively ignore information from it, if the integrity monitoring causes too much nuisance.

**4.2. UAS – Unmanned Aircraft Systems**

Considering the ICAO principles, UAS will operate in accordance with ICAO Standards that exist for manned aircraft as well as any special and specific standards that address the operational, legal and safety differences between manned and unmanned aircraft operations. A few years ago, ICAO [37] stated that, in order for UAS's to be integrated into non-segregated airspace, there shall be a pilot responsible for the UAS operation, however this recommendation already seems outdated and there are serious initiatives to increase airborne autonomy [38]. Anyway, the safe integration of UAS into non-segregated airspace will be a long-term activity with many stakeholders adding their expertise on such diverse topics and needs the development of a robust regulatory framework.

Many businesses involved with the development of small UAS tend to have a short design-to-production cycle, after all quick returns are attractive. Indeed, these businesses have talent in very sophisticated technologies, but often they lack aviation experience. This lack of domain expertise can generate wrong attitudes towards aviation safety and the necessary processes to prove it.

On the other hand, regulations should not be meant to prevent technological advances such as in the UAS sector, either intentionally or unintentionally. EASA has the publication of A-NPA (Advance Notice of Proposed Amendment) 2015-10 [39] proposing an operation regulatory framework for the operation of unmanned aircraft, recognizing how diverse, innovative and international this industry is. It also shows how important it is to identify and implement common rules to promote the safe and secure operations of UAS in all European Member States. EASA has created two task forces to investigate the issues related with the operation of small UAS and the future regulation of the open category. One of these task forces is dedicated to geo-limitation, focusing on the risk of conflict of these UAS with other airspace users. Another task force is assessing the consequences of UAS-Aircraft collisions on the aircraft and its occupants.

The Joint Authorities for Rulemaking on Unmanned Systems (JARUS) Working Groups (WG's) purpose is to recommend a single set of technical, safety and operational requirements for the certification and safe integration of UAS into airspace and it intends to contribute to rulemaking efforts (regional and worldwide). These groups aims at proposing harmonized regulation to cover all aspects of UAS operations considering Operational, Personnel, Technical, and Organizations. JARUS WG-6, on Safety & Risk Assessment, developed a guidance material [40] associated with showing compliance with system safety assessment requirements used in FAA certification (FAA Advisory Circular 23.1309 [41]). The FAA Advisory Circular 23 was not developed for Remotely Piloted Aircraft Systems (RPAS), and does not fully reflect the characteristics of this kind of aircraft, but JARUS WG-6 understands that it is important to maintain the same base of manned aircraft safety assessment. Thus, the document [40] has been produced to provide additional means, but not the only ones, that can be used for showing compliance with the availability and integrity requirements for RPAS systems. It has to be used in conjunction with existing guidance material and to supplement the engineering and operational judgment that should form the basis of any compliance demonstration. The methodology developed in this Acceptable Means of Compliance is based on the objective that RPAS operations must be as safe as manned aircraft. They should not present a hazard to persons or property on the ground or in the air that is any greater than that attributable to the operation of manned aircraft of equivalent class or category. Furthermore, it is assumed that UAS will operate in accordance with the rules governing the flight of manned aircraft and must meet equipment requirements applicable to the class of airspace within which they intend to operate.

The ICAO Circular, EASA studies and JARUS workgroup examples presented in this paper are a small part of the whole UAS certification efforts. It is interesting to observe that the ICAO Member States are investing in working groups that are exchanging experiences once the challenge of UAS integrated into non-segregated airspace are the same for all members. We can observe that there is no solid regulation that could be used internationally, and local national solutions have been presented instead. Our understanding about this observation is that the responsibility over the airspace demands very high safety standards and much more needs to be done in safety methodology to guarantee that the UAS will maintain the actual safety aviation level.

### 4.3. Aircraft Self-Separation

Significant research efforts have been invested in an attempt to make self-organizing air traffic possible, to the point where it was being considered a serious future scenario by the industry about two decades ago [42], a concept known as "Free-Flight". Hoekstra and Ruigrok [43] conducted exploratory studies in collaboration with NASA, and presented arguments defending Self-Separation as safer than the existing ATC system and methods, albeit nobody could soundly demonstrate that Self-Separation is more efficient. The methods for conflict resolution proposed by Hoekstra and Ruigrok do not require cooperation among the aircraft, just knowledge of each other positions.

Traffic Collision Avoidance System (TCAS) has been around for several decades as an autonomous means of avoiding airborne collisions, but it does not have the amount of anticipation and flexibility to allow efficient maneuvers and to be the sole means of separation. The occurrence of a mere TCAS traffic advisory requires a safety report and, because of these reasons, TCAS must not be relied upon as an ordinary means of air traffic management.

After the initial rush for Free-Flight, the research community saw the full breadth and depth of the costs it implies and backed up a little bit. On the other hand, with the affirmation of Automatic Dependent Surveillance (ADS-B), many aviation researchers saw in it a way to implement Free Flight concepts. What resulted from this two-way tendency was hybrid versions of Free-Flight, studied in research projects such as NASA's Distributed Air-Ground Traffic Management (DAG-TM) [44], maintaining the significant role of ATC, but allowing more freedom in en-route airspace.

Vilaplana Ruiz [45] used ADS-B as communication infrastructure and proposed some cooperative solutions for the problem of conflict resolution in a Free-Flight environment, thus allowing more efficient solutions than that of Hoekstra and Ruigrok. The concept of operations was denominated Autonomous Aircraft Operations (AAO), which uses principles of distributed artificial intelligence, a method area also known as Multi-Agent Systems. It requires a cooperation protocol and can provide collective/balanced optimization among the agents and involves team formation and, within the cooperative team, the election of a team organizer for each conflict resolution.

Other aspects relevant to Free Flight have been studied in the Mediterranean Free Flight project (MFF), which concluded that self-separation would be overall beneficial, but it should be limited to low- or medium-density airspace [46]. One of the legacies of this project was the proposal of airspace configurations using Free-Flight sectors where the air traffic density is lower, adjacent to ATC sectors where the traffic density is higher. This gave more breadth to Free Flight and generated a new round of studies under the denomination of Advanced Autonomous Aircraft (A3), which can accommodate high density traffic and is limited mostly by the reliability of ADS-B information and the accuracy of the wind predictions [47]. Within the A3 concept of operations, several authors developed conflict resolution techniques based on Model Predictive Control (MPC), as it can be found in the iFly project documentation [48]. These methods have criteria to minimize deviation from the intended trajectories, however its main limitations are: i) fuel consumption is not part of the model; ii) they require mutually known performance models; iii) they require that every aircraft uses the same control algorithm, otherwise the result may be very inefficient; and iv) they don't have criteria for defining and managing aircraft priority.

## 5. CONCLUSIONS

Systems such as aviation and aerospace are becoming more complex. As a new challenge, the presence of more integrated and more autonomous aviation technologies increases the complexity of such systems, which may exhibit potentially undesirable failure modes. It is important to know the causes of accidents in complex systems in order to improve their safety and to develop preventive strategies to mitigate the occurrence of future accidents.

There is a need for methodologies beyond the ones mandated by the industry, including guidelines for developing the system architecture and control models. Some advances have been made in both argumentative and quantitative analysis methods which allow designers and independent analysts to handle complex and emergent behaviors, however it is still not easy that regulatory authorities accept such new methods for demonstration towards certification or even to conditional approvals.

The complexity of next generation systems offers a challenging area of multidisciplinary research in the development of new safety analysis involving researchers from engineering, social sciences, and cognitive psychology. Thus, there is an urgent need for researchers to look outside their traditional areas in order to capture the complexity of contemporary systems from a systemic view in order to understand the multi-dimensional characteristics of safety. One of the desired outcomes of safety analysis is helping the definitions of constraints to the design, development, and operation of systems with innovative technologies, which prevent later corrective actions. The evaluation of these constraints and other conclusions of safety analysis will be used positively in downstream development stages, first in simulations where next generation aviation concepts would be represented in virtual environments, and later on in physical prototypes.

## 6. ACKNOWLEDGEMENTS

This paper is the result of a joint collaborative work between Boeing Research & Technology and the University of São Paulo, as part of an overarching collaboration agreement between these two organizations. We would like to thank the management and support persons of both institutions which were responsible for turning this possible and those who currently work to maintain the project.